\documentclass{article}

\usepackage{graphicx}%
\usepackage{multirow}%
\usepackage{amsmath,amssymb,amsfonts}%
\usepackage{amsthm}%
\usepackage{mathrsfs}%
\usepackage[title]{appendix}%
\usepackage{xcolor}%
\usepackage{textcomp}%
\usepackage{manyfoot}%
\usepackage{booktabs}%
\usepackage{algorithm}%
\usepackage{algorithmicx}%
\usepackage{algpseudocode}%
\usepackage{listings}%

\usepackage{physics}

\usepackage{mathtools}
\usepackage{color}
\DeclarePairedDelimiter\ceil{\lceil}{\rceil}

\usepackage{nameref}

\usepackage{url}

\usepackage{bibunits}
\usepackage{lineno}

\usepackage{ulem}
\usepackage{authblk}

\title{Tensor-based quantum phase difference estimation for large-scale demonstration}

\date{}
\author[1,8]{Shu Kanno\thanks{shu.kanno@quantum.keio.ac.jp}}
\author[2,3,4,8]{Kenji Sugisaki}
\author[8]{Hajime Nakamura}
\author[5,8]{Hiroshi Yamauchi}
\author[6,8]{Rei Sakuma}
\author[1]{Takao Kobayashi}
\author[1,8]{Qi Gao}
\author[4,7,8]{Naoki Yamamoto}

\affil[1]{Mitsubishi Chemical Corporation, Science \& Innovation Center, Yokohama, 227-8502, Japan}
\affil[2]{Graduate School of Science and Technology, Keio University, 7-1 Shinkawasaki, Saiwai-ku, Kawasaki, Kanagawa 212-0032, Japan}
\affil[3]{Centre for Quantum Engineering, Research and Education, TCG Centres for Research and Education in Science and Technology, Sector V, Salt Lake, Kolkata 700091, India}
\affil[4]{Keio University Sustainable Quantum Artificial Intelligence Center (KSQAIC), Keio University, 2-15-45 Mita, Minato-ku, Tokyo, Japan}
\affil[5]{SoftBank Corp., Research Institute of Advanced Technology, Tokyo, 105-7529, Japan}
\affil[6]{Materials Informatics Initiative, RD Technology \& Digital Transformation Center, JSR Corporation, 3-103-9 Tonomachi, Kawasaki-ku, Kawasaki, 210-0821, Japan}
\affil[7]{Department of Applied Physics and Physico-Informatics, Keio University, 3-14-1 Hiyoshi, Kohoku-ku, Yokohama 223-8522, Japan}
\affil[8]{Quantum Computing Center, Keio University, 3-14-1 Hiyoshi, Kohoku-ku, Yokohama, 223-8522, Japan}

\begin{document}
\maketitle
\begin{abstract}
We develop an energy calculation algorithm leveraging quantum phase difference estimation (QPDE) scheme and a tensor-network-based unitary compression method in the preparation of superposition states and time-evolution gates. 
Alongside its efficient implementation, this algorithm reduces depolarization noise affections exponentially.
We demonstrated energy gap calculations for one-dimensional Hubbard models on IBM superconducting devices using circuits up to 32-system (plus one-ancilla) qubits, a five-fold increase over previous QPE demonstrations, at the 7242 controlled-Z gate level of standard transpilation, utilizing a Q-CTRL error suppression module.
Additionally, we propose a technique towards molecular executions using spatial orbital localization and index sorting, verified linear polyene simulations up to 21 qubits. 
Since QPDE can handle the same objectives as QPE, our algorithm represents a leap forward in quantum computing on real devices.
\end{abstract}

\section{Introduction}
Abilities to calculate physical properties of materials with high accuracy are crucial for accelerating novel material discovery. 
The physical properties are mainly governed by the behavior of electrons in the material, while the number of possible electronic state configurations scales exponentially with system size.
Approximation methods such as the Hartree--Fock (HF) method and density functional theory (DFT) are commonly used to cope with this exponential increase in classical computers. However, these methods often fail when electronic correlations are strong, leading to computational inaccuracies~\cite{Bauer2020-ug, Cao2019-qa, Cohen2012-ji}.
Quantum computers are expected to overcome the accuracy limitations of classical computers because complex quantum states intractable by classical means can be efficiently represented using quantum superposition. 
\label{sec: introduction}

\begin{figure}[h!]
 \centering
 \includegraphics[width=1\textwidth]{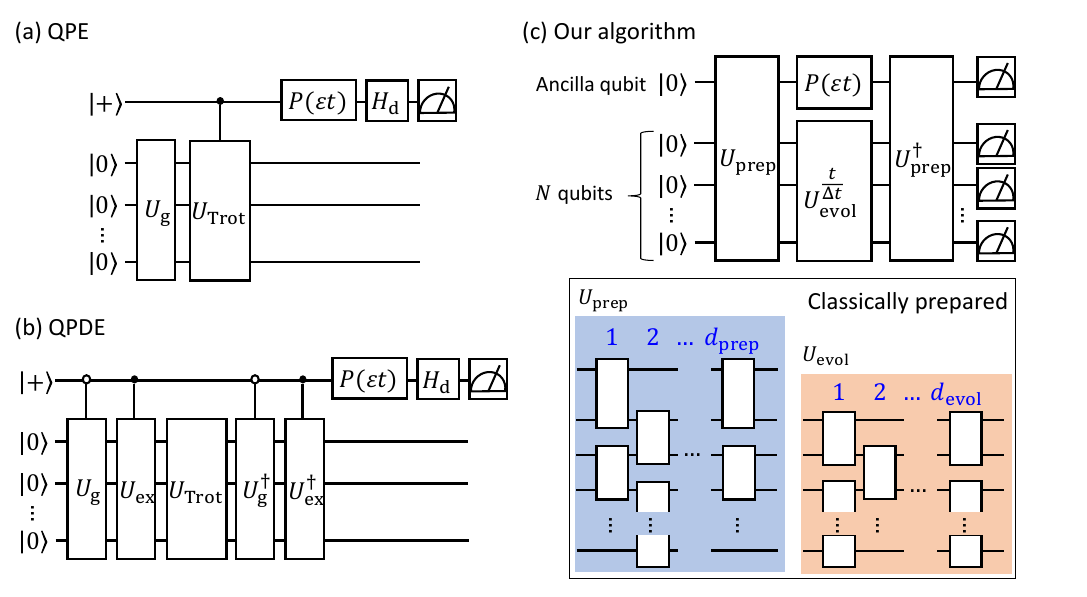}
\caption{Circuits for the Bayesian QPE-type algorithms and our algorithm. In all the figures, the topmost qubit is an ancilla qubit and the others are system qubits. $P(x)$ is a phase gate, and $H_\mathrm{d}$ is a Hadamard gate.
$U_{\mathrm{Trot}}$ is a gate for the time evolution. $U_{\mathrm{g}}$ and $U_{\mathrm{ex}}$ are gates for approximate ground and excited state preparations, respectively.
$U_{\mathrm{prep}}$ is an MPO for approximately preparing the superposition of ground and excited states, and $U_{\mathrm{evol}}$ is an MPO for approximating the time evolution operator. (a) Quantum phase estimation. (b) Quantum phase difference estimation. (c) Our algorithm. Note that while we can classically prepare the compressed operator of a single time step $U_{\mathrm{evol}}$, it becomes intractable to classically simulate total time duration $U_{\mathrm{evol}}^{t/\Delta t}$.} 
 \label{fig: Overview_3column.pdf}
\end{figure}

One of the most important quantum algorithms is quantum phase estimation (QPE)~\cite{Yu_Kitaev1995-cf, Nielsen2010-hw}. 
QPE aims to find the eigenvalue corresponding to a given eigenstate of a system, representing the energy in quantum chemistry. 
In the conventional QPE~\cite{Nielsen2010-hw}, even if the approximate wave function $\ket{\varphi}$ is used as an initial wave function, the full configuration interaction (FCI) energy can be obtained with probability proportional to the squared overlap with the eigenfunction of the target Hamiltonian $H$.
Thus, QPE can potentially compute FCI energies exponentially faster than classical methods, given appropriately prepared $\ket{\varphi}$. 

We focus on Bayesian QPE~\cite{Wiebe2016-kr}, which utilizes Bayesian inference to iteratively narrow the confidence interval of the target phase value. 
The schematic is shown in Fig.~\ref{fig: Overview_3column.pdf}(a); the probability of obtaining 0 in the measurement on the ancilla qubit constitutes a cosine curve for an energy parameter $\varepsilon$, the peak of which approximates the target phase. 
Controlled operations of time evolution gate $U_{\mathrm{Trot}}$ often require significant gate overhead in Bayesian (and conventional) QPE. For instance, a recent experiment for a two-qubit hydrogen molecule simulation on a trapped-ion device used 920 two-qubit gates~\cite{Yamamoto2024-wf}.
Moreover, most of the QPE studies performed one- or two-qubit systems, and even using classical preprocess with exponential cost, a six-qubit system was the maximum~\cite{Lanyon2010-jh, Du2010-jn, Wang2015-um, OMalley2016-hd, Paesani2017-um, Santagati2018-hc, Blunt2023-jf}.
The overhead is particularly challenging in nearest-neighbor architectures like superconducting devices, which require qubit swaps for long-range operations. 
Hence, the quest for gate-efficient implementations is essential for large-scale QPE-type algorithms.

In this work, we propose a gate-efficient, QPE-type energy estimation algorithm.
Our algorithm is based on the quantum phase difference estimation (QPDE)~\cite{Sugisaki2021-gg}, a variant of QPE.
QPDE can also compute FCI energy~\cite{Sugisaki2021-nk} but primarily targets the energy gap between states (e.g., ground and excited states)~\cite{Matsuzaki2020-nx, Sugisaki2021-gg, Russo2021-va}. See Supporting Information~\ref{sec: Procedure of the algorithm} for details.
The important feature of QPDE is an avoidance of costly controlled-$U_{\mathrm{Trot}}$ operations. Instead, as shown in Fig.~\ref{fig: Overview_3column.pdf}(b), it applies controlled operations to $U_{\mathrm{g}}$ and $U_{\mathrm{ex}}$ for preparing superpositions of ground and excited states, where $U_{\mathrm{g}}$ and $U_{\mathrm{ex}}$ are gates for approximate ground and excited state preparations, respectively. 
Our algorithm is a gate-efficient realization of QPDE; specifically, we classically prepare compressed versions of $U_{\mathrm{Trot}}$ as well as the controlled-$U_{\mathrm{g}}$ and $U_{\mathrm{ex}}$ as tensor networks, particularly matrix product operators (MPOs)~\cite{Rudolph2022-ol,Ran2020-li, Lin2021-km,Dborin2022-mh,Dborin2022-sx,Mc_Keever2023-bu,Anselme_Martin2024-dw,Causer2024-wd,Shirakawa2024-yi,  Mc_Keever2024-ji}. 
These classically prepared operators are realized on a quantum circuit shown in Fig.~\ref{fig: Overview_3column.pdf}(c). 
The proposed algorithm offers four main advantages.
First, the resulting MPO-based circuits are constructed with nearest-neighbor gates in a brick-wall layout, facilitating parallelization without swaps. 
Second, the MPO can be efficiently prepared on classical computers and implemented on a quantum circuit, meaning that our algorithm does not suffer from a large amount of quantum-classical communication as in the variational quantum eigensolver~\cite{Peruzzo2014-kp}. 
Third, because the initial superposition state is efficiently prepared as a matrix product state (MPS), the corresponding state preparation gate can be well approximated by an MPO. 
Finally, depolarizing noise effects are exponentially suppressed with respect to the number of qubits. 
The theoretical details can be seen in Supporting Information~\ref{sec: Unitary compression} for the first three points and Supporting Information~\ref{sec: Error analysis for the algorithm} for the final point.

We apply our algorithm to calculate the energy gap for the one-dimensional Hubbard model, utilizing a quantum circuit with up to 32 qubits (plus one ancilla) on an IBM superconducting device, aided by a Q-CTRL error suppression module, where the overview of this module is described on Supporting Information~\ref{sec: Procedure of the algorithm}. Notably, this scale is five times larger than previous QPE studies~\cite{Blunt2023-jf}.
In addition, we investigate linear polyenes of 1,3,5-hexatriene, 1,3,5,7-octatetraene, and 1,3,5,7,9 decapentaene (hereafter referred to as hexatriene, octatetraene, and decapentaene, respectively) as model systems of linear $\pi$-conjugated polyenes. In linear polyenes, the character of the lowest excited singlet state is known to change depending on the length of $\pi$-conjugation~\cite{Nakayama1998-vl}, and the chemistry of excited states is of significant importance.
In the calculation of hexatriene, a one-dimensional 12-qubit Hamiltonian was constructed by using newly developed orbital localization and index reordering techniques. 
We also execute it on octatetraene and decapentaene of 16- and 20-qubit Hamiltonian, respectively, in Supporting Information~\ref{sec: Error analysis for the algorithm}.
Finally, recall that the MPO-based time-evolution gate of short time $\Delta t$ can be classically implemented, while its concatenation for long-time simulation cannot. Thus, the quantum advantage of this algorithm lies in the time evolution circuit. See Supporting Information~\ref{sec: time scale for the quantum advantage} for the numerical verification of exponential growth for bond dimensions in an MPO.

\section{Tensor-based phase difference estimation algorithm}

The entire estimation algorithm is summarized to Algorithm~\ref{Alg: TQPDE} given in Supporting Information~\ref{sec: Procedure of the algorithm}, and here we describe the procedure sketch. 
For each iteration, the multiple circuits with different parameters $\varepsilon$ determined from the prior distribution are executed, where the prior and posterior distributions are assumed to be Gaussian, and the total time $t$ is determined from the variance of the prior distribution. 
In advance to executing the quantum algorithm with the circuit Fig.~\ref{fig: Overview_3column.pdf}(c), we compute MPOs corresponding to $U_{\mathrm{prep}}$ and $U_{\mathrm{evol}}$ by classical means, followed by preparing their gate realization. 

The quantum algorithm is executed as follows;
we prepare the superposition state using $U_{\mathrm{prep}}$ gate, evolve the state by operating $U_{\mathrm{evol}}$ by $t/\Delta t$ times, operate the phase gate with $\varepsilon t$, uncompute the state using $U_{\mathrm{prep}}^{\dagger}$ gate, and measure all qubits to obtain the probability of getting all 0. 
The number of depths of $U_{\mathrm{prep}}$ and $U_{\mathrm{evol}}$ are $d_{\mathrm{prep}}$ and $d_{\mathrm{evol}}$, respectively. 
We repeat this procedure with different $\varepsilon$ to obtain the likelihood function which can be further approximated by Gaussian distribution. 
The posterior distribution is calculated to update the confidence interval of $\varepsilon$, and the posterior distribution is used as the prior distribution for the next iteration.

We can also calculate the FCI energy by simplifying the procedure. 
That is, the value of gap with a negative sign corresponds to the FCI energy when $\ket{\psi_{\mathrm{ex}}}$ is a vacuum state~\cite{Sugisaki2021-nk}, and see Supporting Information~\ref{sec: FCI energy calculation} for the demonstration of the FCI energy calculation.
Note that by extending the tensor structure from an MPO to a complicated tensor such as a tree tensor, our algorithm would be useful even for all-to-all connected devices such as trapped-ion and neutral atom devices since the long-distance transport of ions (or atoms) is costly in practice~\cite{Brown2016-be}.

\section{Benchmark on the unitary compression method}
\label{sec: Brief comparison with the conventional method}

We provide a brief comparison of the proposed unitary compression method with conventional methods for state preparation and time evolution circuits.
We consider the following one-dimensional Hubbard  model:
\begin{equation}
\begin{aligned}
H &= -T \sum_{q=1}^{n_{\mathrm{s}}-1} \sum_{\sigma \in \{\uparrow, \downarrow \}}(a_{q+1\sigma}^{\dagger} a_{q\sigma}+a_{q\sigma}^{\dagger} a_{q+1\sigma})
+U \sum_{q=1}^{n_{\mathrm{s}}} n_{q\uparrow} n_{q\downarrow}
-\frac{U}{2} \sum_{q=1}^{n_{\mathrm{s}}} (n_{q\uparrow}+n_{q\downarrow}),
\label{Eq: Hubbard model}
\end{aligned}
\end{equation}
where $q$ ($\sigma$) is the orbital (spin) index, $T=1$ is the hopping energy so that energy is unitless, $U$ is the on-site Coulomb energy, $a_{q\sigma}^{\dagger}$ ($a_{q\sigma}$) is the creation (annihilation) operator, and $n_{q\sigma}$ is the number operator $n_{q\sigma} = a_{q\sigma}^{\dag} a_{q\sigma}$. 
In this study, we choose $U=10$, which corresponds to the strong correlation regime, and we consider $n_{\mathrm{s}} = 4$, i.e., eight-system qubit model in this section, and the fermion-qubit mapping is a Jordan--Wigner transformation with an up-down qubit sequence $1\uparrow, 1\downarrow, 2\uparrow,\dots,n_{\mathrm{s}}\downarrow$.

The metric to evaluate the $U_{\mathrm{prep}}$ approximation for the state preparation circuit is $f(U_{\mathrm{prep}};~\ket{\mathrm{MPS}})$ represented by
\begin{equation}
\begin{aligned}
f(U_{\mathrm{prep}};~\ket{\mathrm{MPS}}) = \bra{\mathrm{MPS}} U_{\mathrm{prep}} \ket{0}^{\otimes N+1},
\label{Eq: prep metric}
\end{aligned}
\end{equation}
where $N$ is the number of system qubits and $N=2n_{\mathrm{s}}$ in the Hubbard model. 
Because we consider only the real wave function in this study, the metric is real. 
$\ket{\mathrm{MPS}}$ is an MPS calculated for the ground and first-excited states using the density matrix renormalization group (DMRG)~\cite{White1992-dk,Schollwock2011-im}. 
Because we confirmed that the DMRG result was almost the same as the exact solution in the current eight-qubit model, $f(U_{\mathrm{prep}};~\ket{\mathrm{MPS}})$ is close to the square root of fidelity $f(U_{\mathrm{prep}};~\ket{\psi_{\mathrm{target}}})$, where $\ket{\psi_{\mathrm{target}}} = \frac{1}{\sqrt{2}}(\ket{0}\ket{\psi_{\mathrm{g}}}+\ket{1}\ket{\psi_{\mathrm{ex}}})$, and $\ket{\psi_{\mathrm{g}}}$ and $\ket{\psi_{\mathrm{ex}}}$ are approximate ground and excited states obtained using the DMRG, respectively.
The metric $f(U_{\mathrm{prep}};~\ket{\mathrm{MPS}})$ takes the maximum value 1 when $U_{\mathrm{prep}} \ket{0}^{\otimes N+1}=\ket{\mathrm{MPS}}$.
The coefficient of the Hartree--Fock configuration in the exact ground state is 0.66. 
On the other hand, the $f(U_{\mathrm{prep}};~\ket{\mathrm{MPS}})$ for this model was 0.99 at $d_{\mathrm{prep}}=6$,
which indicates that the $U_{\mathrm{prep}}$ can prepare states with high accuracy.

\begin{figure}[]
 \centering
 \includegraphics[width=0.5\textwidth]{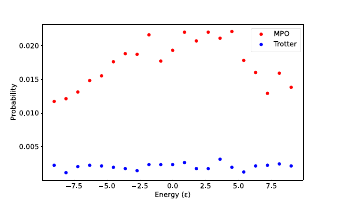}
\caption{Comparison of the time evolution circuits (nine qubits) of the first-order Trotter approximation and that of the MPO. The Eagle device $ibm\_osaka$ was adopted. The probability of all 0 measurements for each of the phase $\varepsilon$ is plotted.}
 \label{fig: Compare_1column.pdf}
\end{figure}

The error metric of the $U_{\mathrm{evol}}$ approximation in the time evolution circuit is $\delta(U_{\mathrm{evol}};~U_{\mathrm{ref}})$ as in Ref.~\cite{Causer2024-wd}:
\begin{equation}
\begin{aligned}
\delta(U_{\mathrm{evol}};~U_{\mathrm{ref}}) = \sqrt{2-\Re\Tr[U_{\mathrm{ref}}^{\dagger}U_{\mathrm{evol}}]^{\frac{1}{N}}},
\label{Eq: evol metric}
\end{aligned}
\end{equation}
where $U_{\mathrm{ref}}$ is a reference MPO prepared by converting the time evolution operator to an MPO (see Supporting Information~\ref{sec: Unitary compression} for details). 
$\frac{1}{N}$ is the normalization factor.
From the related reason in the state preparation of $\ket{\mathrm{MPS}} \approx \ket{\psi_{\mathrm{target}}}$, $\delta(U_{\mathrm{evol}};~U_{\mathrm{ref}})$ is close to $\delta(U_{\mathrm{evol}};~e^{-iH\Delta t})$ in the current model.
We choose $\Delta t = 0.1$ throughout this study.
$\delta(U_{\mathrm{evol}};~U_{\mathrm{ref}})$ takes the minimum value 0 when $U_{\mathrm{evol}} = U_{\mathrm{ref}}$.
$\delta(U_{\mathrm{evol}};~U_{\mathrm{ref}})$ at $d_{\mathrm{evol}} = 5$ was $4.3\times10^{-3}$, and $\delta(U_{\mathrm{evol}};~e^{-iH\Delta t})$ of the first- and second-order Trotter approximations were $2.2\times10^{-2}$ and $1.6\times10^{-3}$, respectively.
The values suggest that the accuracy of prepared $U_{\mathrm{evol}}$ is between first- and second-order Trotter approximations, which corresponds to the tendency in the previous study~\cite{Causer2024-wd}.

To confirm the circuit execution efficiency, we compared the results of the measured probability distributions in the circuit of Fig.~\ref{fig: Overview_3column.pdf}(c) run on a real device with $U_{\mathrm{evol}}$ as is and replaced by the first-order Trotter approximation. 
The variance of the prior distribution is $9.0$ which corresponds to $t=0.2$ (see Supporting Information~\ref{sec: Procedure of the algorithm}), the device is Eagle $ibm\_osaka$, $d_{\mathrm{prep}}=6$, and the error suppression module appeared in the subsequent sections is not used here.
The number of shots is 10,000 in the real device execution of this study.
The number of two-qubit gates in the circuit after transpilation in Qiskit~\cite{Javadi-Abhari2024-qm} was 216 for $U_{\mathrm{prep}}$ and 234 for the first-order Trotter approximation.
Here, we consider the two-qubit gates as the ECR gate and the controlled-Z gate in the Eagle and Heron devices, respectively.
The results are shown in Fig.~\ref{fig: Compare_1column.pdf}. 
The plots by $U_{\mathrm{evol}}$ (red) show peaks, while the Trotter results (blue) show a uniform distribution.
These results indicate that the unitary compression has a high execution efficiency for real devices.

\begin{figure}[t]
 \centering
 \includegraphics[width=1\textwidth]{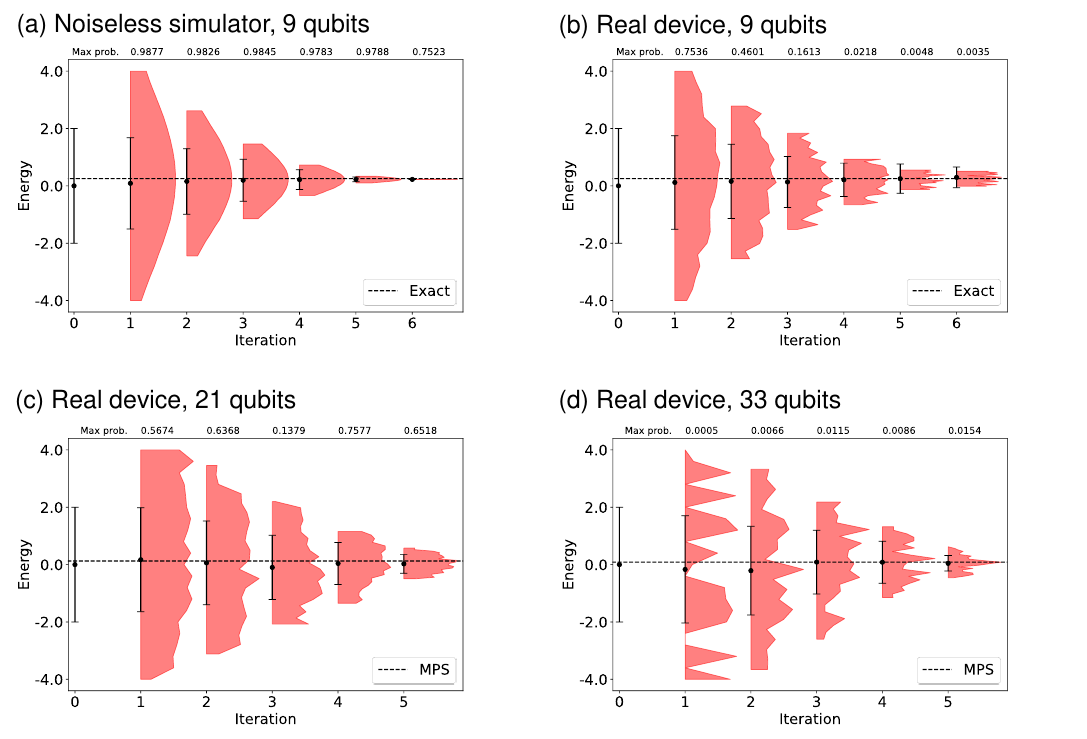}
\caption{Demonstration of our algorithm in the one-dimensional Hubbard models using (a) the noiseless simulator and (b,c,d) the Heron device $ibm\_torino$ with the error suppression. The circle and error bar denote the mean value and standard deviation, respectively, of the posterior distribution in each iteration. The sampled probabilities are shown in red, and each plot is normalized by the maximum value in each iteration, which is shown at the top of the plot.}
 \label{fig: Demonstration_3column.pdf}
\end{figure}

\section{Demonstration of the algorithm}
\label{sec: Demonstration of our algorithm}
We show the demonstration for the Hubbard model in~\eqref{Eq: Hubbard model}.
Figure~\ref{fig: Demonstration_3column.pdf}(a) shows the result of a noiseless simulator for our algorithm with $n_{\mathrm{s}}=4$, i.e., a nine-qubit circuit including one ancilla qubit, where $d_{\mathrm{prep}}=6$ and $d_{\mathrm{evol}}=5$ which are the same setting in the previous section.
As the iteration increases, the mean value of the posterior distribution (circle) approaches the exact energy (dashed line), and the standard deviation (error bar) becomes smaller.
The mean value of the posterior distribution in the final iteration is $0.224\pm0.005$, and the exact value is $0.254$; i.e., the bias error $0.030$ remains. 
From the results of $\delta(U_{\mathrm{evol}};~U_{\mathrm{ref}})$ in the previous section, the accuracy is improved by increasing  $d_{\mathrm{evol}}$ from five; in fact, if $d_{\mathrm{evol}}=8$ and 10, the bias error decreases to 0.020 and 0.012, respectively.
The probabilities taken from each circuit with $\varepsilon$ determined from the prior distribution are shown in red, and a peak at the mean value can be seen.

Figure~\ref{fig: Demonstration_3column.pdf}(b,c,d) show the results on the real device $ibm\_torino$, where we used an error suppression module FireOpal in Q-CTRL~\cite{Mundada2023-rl,Sachdeva2024-vr,Baum2021-sd}. 
Certainly in Fig.~\ref{fig: Demonstration_3column.pdf}(b), the maximum probability of 0.75 at the first iteration was reduced to 0.0035 at the sixth iteration due to noise.
Nevertheless, it was observed that the mean values move toward the exact solution, and the variance also decreases as the iterations increase. That is, our algorithm worked in the nine-qubit circuit, where the final value was $0.294\pm0.359$.
The two-qubit gate counts in the circuit in the final iteration is 2984/2274, where hereafter we regard the gate count value before and after slash as the value before and after the error suppression, respectively. 
Note that the error suppression was executed after the Qiskit transpilation.

Figures~\ref{fig: Demonstration_3column.pdf}(c) and (d) are results for $n_{\mathrm{s}}=10$ and $16$ by using 21 and 33 qubit circuits, where  $f(U_{\mathrm{prep}};~\ket{\mathrm{MPS}})$ in $d_{\mathrm{evol}}=12$ are 0.92 and 0.81, respectively.
$d_{\mathrm{prep}}$ was fixed to five since the values $\delta(U_{\mathrm{evol}};~U_{\mathrm{ref}})$ tend not to depend on the model size, where the values are $4.6\times10^{-3}$ in 20 system qubits, $4.5\times10^{-3}$ in 32 qubits, and $5.3\times10^{-3}$ even in 100 qubits.
Although there are some numerical instabilities, such as in the low probability in the first iteration in Fig.~\ref{fig: Demonstration_3column.pdf}(d), we still obtain final gap values of $0.026\pm0.330$ and $0.043\pm0.270$ for 21 and 33 qubit circuits, and the MPS gap values for references are 0.125 and 0.084, respectively.
As in Supporting Information~\ref{sec: Error analysis for the algorithm}, we theoretically confirm that our algorithm does not change the peak top value if a depolarization noise is assumed, which is the same as in the previous Bayesian QPE algorithms~\cite{Wiebe2016-kr, Sugisaki2021-nk}.
Furthermore, the effect of depolarization noise in a circuit execution exponentially reduces for $N$ (with a fixed error rate). 
In fact, while the distribution for a nine-qubit model becomes blurred due to noise on the final iteration, sharp peaks can be found for 21- and 33-qubit models even though especially in the 33-qubit model, only 1.54\% of the probability amplitude remains.
Additionally, the number of two-qubit gates in the final iteration of the 33-qubit circuit is 7242/794, nearly 10 times gate reduction by the error suppression was found.
As a result, our algorithm can be performed on more than five times larger systems in terms of the number of qubits than the previous QPE study of six qubit systems~\cite{Blunt2023-jf}.
Note that for the signal weakening in the first iteration on the 33 qubit demonstration, it is difficult for us to fully identify the cause due to the multiple error suppression components in FireOpal (see also Supporting Information~\ref{sec: Procedure of the algorithm}), but it may be relevant that the number of gates in the circuit after the error suppression was slightly high, where the gate counts in the first and final iterations were 819 and 794, respectively. 
We also note that there were only a few signals when the suppression module was not enabled, and the probabilities were too low to obtain meaningful results on the 21-qubit circuit (see Supporting Information~\ref{sec: Error analysis for the algorithm}).

\begin{figure}[]
 \centering
 \includegraphics[width=1\textwidth]{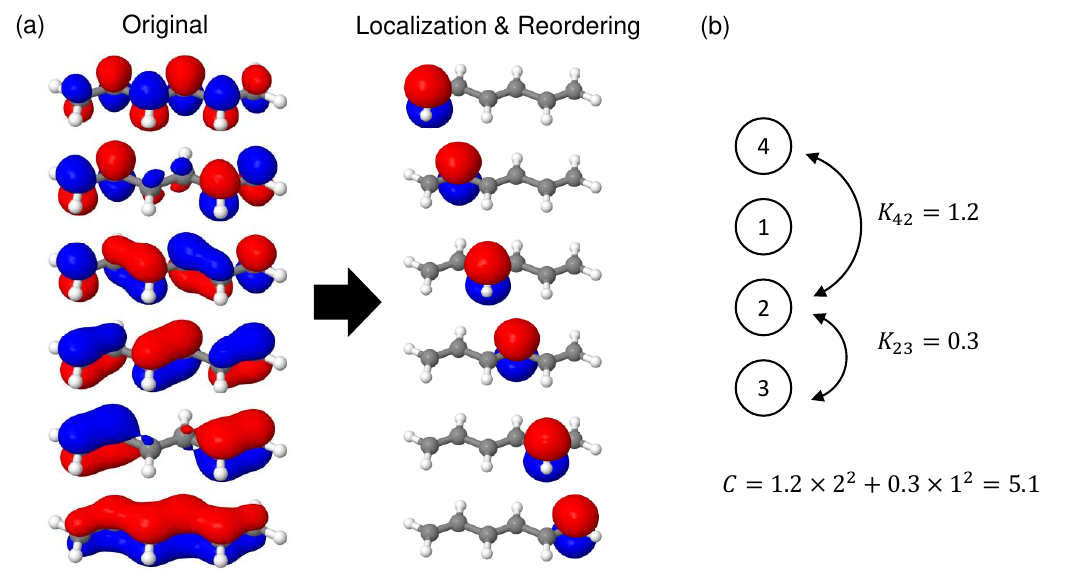}
\caption{Procedure for converting the molecular Hamiltonian to a one-dimensional one. (a) Orbitals before and after the procedure. The structures are drawn by Jmol~\cite{Jmol_development_team2016-ly}. (b) Example of the cost function.}
 \label{fig: Linearization_3column.pdf}
\end{figure}

\begin{figure}[t]
 \centering
 \includegraphics[width=1\textwidth]{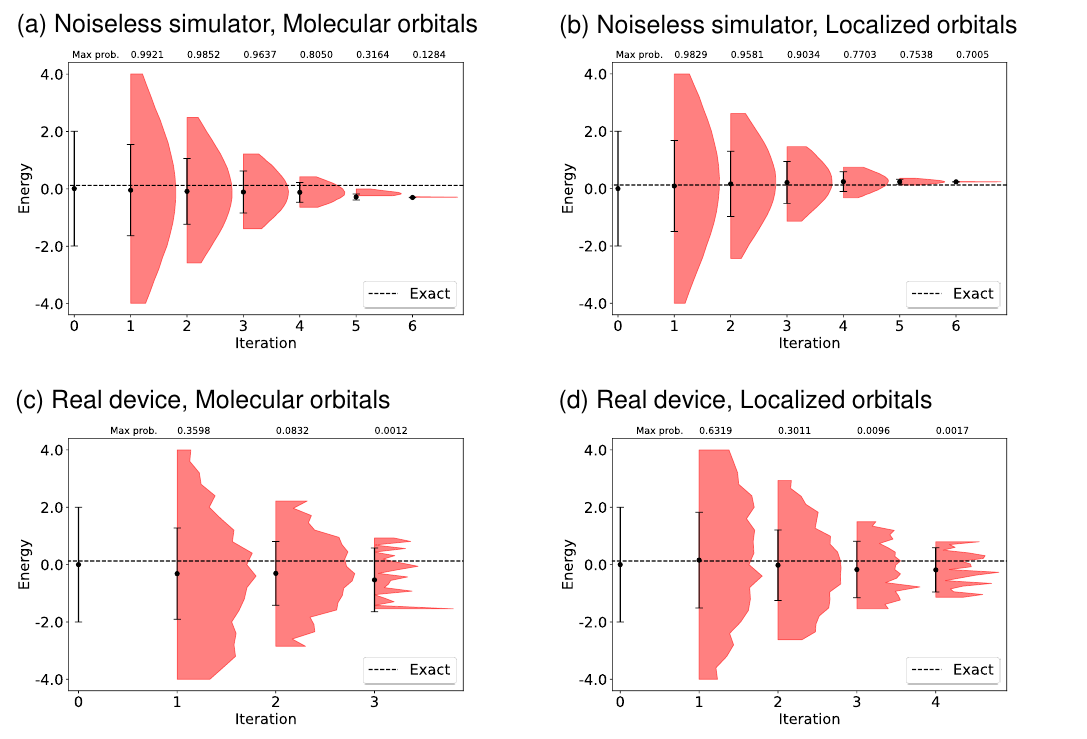}
\caption{Demonstration of our algorithm in hexatriene (13-qubit circuit) using (a,b) the noiseless simulator and (c,d) the Heron device $ibm\_torino$ with the error suppression. The circle and error bar denote the mean value and standard deviation, respectively, of the posterior distribution in the iteration. The sampled probabilities are shown in red, and each plot is normalized by the maximum value in each iteration, which is shown at the top of the plot.}
 \label{fig: Hexatriene_3column.pdf}
\end{figure}

We next show the result of our algorithm to a realistic chemical model.
The target molecule is hexatriene, and the Hamiltonian is a 12-qubit complete active space configuration interaction (CASCI) (6e 6o) problem consisting of $\pi$ and $\pi^*$ orbitals (see Supporting Information~\ref{sec: model construction}).
The obstacle to applying our algorithm in this model is that the molecular orbitals are delocalized throughout the molecule and thus the Hamiltonian is not one-dimensional.
Therefore, we propose a general procedure to convert such Hamiltonian to a one-dimensional one, as shown in Fig.~\ref{fig: Linearization_3column.pdf}(a). 
We first localized the orbitals by Boys localization~\cite{Foster1960-kw}.
Then, the orbital indices are reordered by exchange interaction.
Specifically, we introduce the cost function referencing the proposal in the context of tensor networks in classical calculations~\cite{Nakatani2013-cf}, 
\begin{equation}
\begin{aligned}
    C=\sum_{ij} K_{ij} D_{ij}^2,
    \label{eq: orbital sorting cost}    
\end{aligned}
\end{equation}
where $i$ and $j$ are orbital indices, $K_{ij}$ is the value of the exchange interaction between orbitals, and $D_{ij}$ is the distance between orbitals when the orbital indices are arranged in a one-dimensional list. We show an example of $C$ in Fig.~\ref{fig: Linearization_3column.pdf}(b).
$C$ was optimized with a genetic algorithm (see Supporting Information~\ref{sec: model construction}), and its initial and final value is 0.0754 to 0.0160, respectively.
Finally, the reordered model was mapped to qubits by Jordan--Wigner transformation with the up-down qubit sequence.

We calculate the energy gap between the lowest spin-singlet and triplet states. The values of the metric $f(U_{\mathrm{prep}};~\ket{\mathrm{MPS}})$ in the original and in the one-dimensionalized models in $d_{\mathrm{prep}}=10$ were 0.94 and 0.99, which indicates that $U_{\mathrm{prep}}$ is producing high fidelity despite the fact that both the ground and excited state cannot be approximated by a single configuration due to orbital localization.
The values of $\delta(U_{\mathrm{evol}};~U_{\mathrm{ref}})$ in $d_{\mathrm{evol}}=8$ were $1.6\times10^{-2}$ and $9.4\times10^{-3}$, respectively, more than 40\% reduction in $U_{\mathrm{evol}}$.
Note that although the value of the improvement in the metric itself is not large, the improvement is significant because $U_{\mathrm{evol}}$ is applied $t/\Delta t$ times.

Figure~\ref{fig: Hexatriene_3column.pdf} shows the results of our algorithm for each model.
The gap values in the original and one-dimensional models for the noiseless simulator in Figs.~\ref{fig: Hexatriene_3column.pdf}(a) and (b) are $-0.300\pm0.008$ and $0.236\pm0.005$, and the values for the real device in Figs.~\ref{fig: Hexatriene_3column.pdf}(c) and (d) are $-0.534\pm1.112$ and $-0.185\pm0.769$, and the gate counts are 2190/1054 and 2718/658
, respectively.
The exact gap is 0.125 in this system.
The results suggest that the one-dimensional processing can extend the range of applications of our algorithm. 
Note that we also executed octatetraene (decapentaene) of the localized orbital model on 17 (21) qubit circuits in Supporting Information~\ref{sec: Error analysis for the algorithm}. In octatetraene, we could not prepare $U_{\mathrm{evol}}$ of the molecular orbital model even calculating 300 hours on a high-performance computer having more than 100 CPU cores and 1 TB RAM while that of the localized model can be calculated within two days using less than 20 CPU cores.
This fact also suggests the improvements in our procedure.
Furthermore, in order to compare the accuracy of the interaction complexity, we performed our algorithm on five structures of $\mathrm{H_4}$ clusters that gradually changed from linear to square structures using noiseless simulations, see Supporting Information~\ref{sec: additional benchmarks for interaction complexity}.

\section{Conclusions and outlook}
\label{sec: outlook}
We have proposed a gate-efficient and noise-robust quantum phase estimation-type algorithm based on unitary compression by tensor networks.
We applied the algorithm to a one-dimensional Hubbard model and confirmed its effectiveness on a real superconducting device, utilizing an error suppression module, with circuits up to 33 qubits including one ancilla qubit.
Furthermore, for the calculation of chemical models, we proposed a procedure to convert the model composed of molecular orbitals into a one-dimensional model by orbital localization and reordering, achieving over 40\% improvement in the error metric.
Based on these results, it has been demonstrated that QPE, which had previously been validated only in toy models with a small number of qubits, can be executed at a scale comparable to the classical computational limit of FCI, where the current limit is around 40 qubits, depending on the computing environment~\cite{Sun2020-sa, Gao2024-aw}.

The next step is to tackle tasks that are completely infeasible for classical computation, thereby achieving the so-called quantum advantage.
The most critical challenge is an improvement of estimation to achieve chemical accuracy.
A deeper MPO improves accuracy, but in general, the computational cost increases exponentially with depth.
Since our unitary compression can be extended to tensor networks other than MPO, tensor networks that account for non-linearity in device connectivity --- for example heavy hex on IBM devices~\cite{Kim2023-xg} or two-dimensional grids on Google devices~\cite{Arute2019-hj}. In the case of ion- or atom-trapped hardware, physical constraints on the transport of the particles would influence the optimal tensor structure for the algorithm.
Alternatively, integrating quantum-classical hybrid-optimized circuit to $U_{\mathrm{prep}}$ can improve accuracy~\cite{Dborin2022-mh, Nakaji2024-ud, Peruzzo2014-kp}.
Additionally, exploring the use of partially error-corrected devices anticipated in the near future could further advance the feasibility of our approach.

In the fully error-corrected era, the qubitization~\cite{Low2019-eh} becomes an alternative for approximating time evolution operator because it can realize the time evolution with $\order{\log(\epsilon^{-1})}$ queries in contrast to the Trotter method of $\order{\mathrm{poly}(\epsilon^{-1})}$ depth for accuracy $\epsilon$ while the qubitization is currently impractical due to a very high constant cost of the linear combination of unitaries (LCU).
The classical circuit compression would also be useful even in the era, for example, in the PREPARE channel compression in LCU~\cite{Morisaki2024-ot}. It also should be mentioned that the Trotter method can achieve $\order{\mathrm{poly}\log(\epsilon^{-1})}$ by using algorithmic error mitigations~\cite{Endo2019-oh, Watson2024-bn}.

Finally, in addition to QPE forming the basis of quantum computation, the technique we proposed for superposition state preparation to avoid long-range interactions involving an ancilla qubit can be used for more general circuit compilation~\cite{Kanno2024-oz}.
Hence, the proposed algorithms or techniques have potential applications in other fields besides chemistry, such as the Harrow--Hassidim--Lloyd algorithm~\cite{Harrow2009-hp} in machine learning.

\section{Materials and methods}
The technical details of our algorithm are described in Supporting Information~\ref{sec: Procedure of the algorithm},~\ref{sec: Unitary compression}, and~\ref{sec: model construction}.
Supporting Information~\ref{sec: Procedure of the algorithm} shows the procedure of our phase difference estimation algorithm and calculation conditions.
Supporting Information~\ref{sec: Unitary compression} explains how to compress the initial state and time evolution operator via tensor networks. Supporting Information~\ref{sec: model construction} describes the construction of the Hamiltonians in the Hubbard models and the linear polyenes.

\section{Code and data availability}
The codes and datasets in our study can be available at \url{https://github.com/sk888ks/TQPDE_open.git}.  

\section{Acknowledgements}
This work was supported by Quantum Leap Flagship Program (Grant No. JPMXS0118067285 and No. JPMXS0120319794) from the MEXT, Japan. A part of this work was performed for Council for Science, Technology and Innovation (CSTI), Cross-ministerial Strategic Innovation Promotion Program (SIP), “Promoting the application of advanced quantum technology platforms to social issues”(Funding agency: QST).
The part of calculations was performed on the Mitsubishi Chemical Corporation (MCC) high-performance computer (HPC) system “NAYUTA”, where “NAYUTA” is a nickname for MCC HPC and is not a product or service name of MCC.
We acknowledge the use of IBM Quantum services for experiments in this paper. The views expressed are those of the authors, and do not reflect the official policy or position of IBM or the IBM Quantum team.
This work is partly supported by Q-CTRL.
S.K. thanks to Yuki Sato, Toshinari Itoko, Hiroyoshi Kurogi, Hiroshi Watanabe, Yoshiharu Mori, Takashi Abe, Miho Hatanaka, Tamiya Onodera, and Kaito Wada for fruitful discussions, Hajime Sugiyama for the technical support on HPC, Kimberlee Keithley for assistance in reviewing the English manuscript. 
K.S. acknowledges support from Center of Innovations for Sustainable Quantum AI (JPMJPF2221) from JST, Japan, and Grants-in-Aid for Scientific Research C (21K03407) and for Transformative Research Area B (23H03819).

\appendix
\clearpage

\renewcommand{\thesection}{S\arabic{section}} 
\setcounter{section}{0}

\renewcommand{\thepage}{S\arabic{page}}
\setcounter{page}{1}

\renewcommand{\thefigure}{S\arabic{figure}}
\setcounter{figure}{0}

\renewcommand{\thetable}{S\arabic{table}}
\setcounter{table}{0}

\numberwithin{equation}{section}

\section*{Supporting Information}

\section{Procedure of the algorithm}
\label{sec: Procedure of the algorithm}
Algorithm~\ref{Alg: TQPDE} shows the procedure of our algorithm.
In each iteration, we update the distribution along with the measurement results. 
For $\varepsilon$, $t$, and $\Vec{x}$ (i.e., the computational-basis measurement result represented by a binary sequence),
there is a relation among prior distribution $P(\varepsilon)$, likelihood function $P(\Vec{x}| \varepsilon)$, and posterior distribution $P(\varepsilon|\Vec{x})$ called the Bayes theorem,
\begin{equation}
\begin{aligned}
P(\varepsilon | \Vec{x}) \propto P(\Vec{x}| \varepsilon)P(\varepsilon).
\label{Eq: bayes theorem}
\end{aligned}
\end{equation}
The calculated posterior distribution is used as the prior distribution in the next iteration. 
In this work, we take the Bayes setting with Gaussian probability distribution $P(z)=\mathcal{N}(z; \mu, \sigma^2)$ of the random variable $z$, with mean $\mu$ and variance $\sigma^2$. 
We specifically choose the Gaussian prior $P(\varepsilon) = \mathcal{N}(\varepsilon ; \mu_{\mathrm{prior}}, \sigma^2_{\mathrm{prior}})$. 

This study considers distributions over two patterns for all qubits $\vec{x} \in \{\Vec{0}, \mathrm{others}\}$ where $\Vec{0}$ represents all zero measurements, which corresponds to the two measurement patterns for an ancilla qubit, $\{0, 1\}$, in the previous studies~\cite{Wiebe2016-kr, Sugisaki2019-ch}. 
Then, as shown just later, the likelihood function with respect to $\varepsilon$ can be well approximated by the Gaussian $P(\Vec{0}| \varepsilon) = \mathcal{N}(\varepsilon; \mu_{\mathrm{lh}}, \sigma^2_{\mathrm{lh}})$. 
As a result, the posterior distribution keeps the form of Gaussian as $P(\varepsilon | \Vec{0}) = \mathcal{N}(\varepsilon; \mu_{\mathrm{post}}, \sigma^2_{\mathrm{post}})$, where 
\begin{equation}
\begin{aligned}
\mu_{\mathrm{post}} = \frac{\sigma^2_{\mathrm{prior}}\mu_{\mathrm{lh}}+\sigma^2_{\mathrm{lh}}\mu_{\mathrm{prior}}}{\sigma^2_{\mathrm{lh}}+\sigma^2_{\mathrm{prior}}}, 
\label{Eq: update formula mu}
\end{aligned}
\end{equation}
\begin{equation}
\begin{aligned}
\sigma^2_{\mathrm{post}} = \frac{\sigma^2_{\mathrm{prior}}\sigma^2_{\mathrm{lh}}}{\sigma^2_{\mathrm{lh}}+\sigma^2_{\mathrm{prior}}}.
\label{Eq: update formula var}
\end{aligned}
\end{equation}

\begin{algorithm}
\caption{Tensor-based quantum phase difference estimation}
\label{Alg: TQPDE}
\begin{algorithmic}[1]
\State \textbf{Input:} Initial mean (variance) as a prior Gaussian distribution $\mu_{\mathrm{init}}$ ($\sigma^2_{\mathrm{init}}$), the number of sample points $m$, the number of shots $R$, single time step $\Delta t$

\Comment{Initial preparation}
\State $\mu_{\mathrm{prior}} \gets \mu_{\mathrm{init}}$
\State $\sigma^2_{\mathrm{prior}} \gets \sigma^2_{\mathrm{init}}$
\State $iter \gets 1$

\Comment{Estimation loop}
\While{true}
    \State $t \gets 1.8 / \sigma^2_{\mathrm{prior}}$
    \State $\Vec{\varepsilon}_{\mathrm{list}} \gets$ make a list by selecting $m$ points equally in the interval from $\mu_{\mathrm{prior}} - \sigma^2_{\mathrm{prior}}$ to $\mu_{\mathrm{prior}} + \sigma^2_{\mathrm{prior}}$
    \State $\Vec{p}_{\mathrm{list}} \gets$ empty list
    \ForAll{$\varepsilon$ \textbf{in} $\Vec{\varepsilon}_{\mathrm{list}}$}
        \State Execute the circuit in Fig.~\ref{fig: Overview_3column.pdf}(c) with $t$, $\Delta t$, $\varepsilon$, and shot number $R$
        \State Obtain the probability $p$ that $0$ is measured in all qubits of the circuit
        \State Append the value pair ($\varepsilon, p$) to $\Vec{p}_{\mathrm{list}}$
    \EndFor

    \Comment{Mean and variance of likelihood function}
    \State $\mu_{\mathrm{lh}}, \sigma^2_{\mathrm{lh}} \gets$ calculate from the Gaussian fitting of $\Vec{p}_{\mathrm{list}}$

    \Comment{Mean and variance of posterior distribution}
    \State Calculate $\mu_{\mathrm{post}}, \sigma^2_{\mathrm{post}}$ by Eqs.~\eqref{Eq: update formula mu} and~\eqref{Eq: update formula var} using $\mu_{\mathrm{prior}}, \sigma^2_{\mathrm{prior}}, \mu_{\mathrm{lh}}$, and $\sigma^2_{\mathrm{lh}}$
    \State Print $iter, \Vec{p}_{\mathrm{list}}, \mu_{\mathrm{post}}$, and $\sigma^2_{\mathrm{post}}$

    \If{a termination condition, such as threshold of $\sigma^2_{\mathrm{post}}$, is satisfied}
        \State \textbf{break}
    \Else
        \State $\mu_{\mathrm{prior}} \gets \mu_{\mathrm{post}}$
        \State $\sigma^2_{\mathrm{prior}} \gets \sigma^2_{\mathrm{post}}$
        \State $iter \gets iter + 1$
    \EndIf
\EndWhile
\end{algorithmic}
\end{algorithm}

In the circuit implementation in Fig.~\ref{fig: Overview_3column.pdf}(c), we assume
\begin{equation}
\begin{aligned}
U_{\mathrm{prep}}\ket{0}^{\otimes N+1} \approx \frac{1}{\sqrt{2}}(\ket{0}\ket{\psi_{\mathrm{g}}}+\ket{1}\ket{\psi_{\mathrm{ex}}}),
\label{Eq: uprep approximation}
\end{aligned}
\end{equation}

\begin{equation}
\begin{aligned}
U_{\mathrm{evol}} \approx e^{-iH\Delta t},
\label{Eq: evol approximation}
\end{aligned}
\end{equation}
where $\ket{\psi_{\mathrm{g}}}$ and $\ket{\psi_{\mathrm{ex}}}$ are the ground and excited states, respectively, and $N$ is the number of system qubits.
The wave function after operating $U_{\mathrm{evol}}^{t/\Delta t}$ and $P(\varepsilon t)$ is
\begin{equation}
\begin{aligned}
&(P(\varepsilon t)\otimes U_{\mathrm{evol}}^{t/\Delta t}) \frac{1}{\sqrt{2}}(\ket{0}\ket{\psi_{\mathrm{g}}}+\ket{1}\ket{\psi_{\mathrm{ex}}}) \\
&\approx \frac{1}{\sqrt{2}}(e^{-iE_{\mathrm{g}}t}\ket{0}\ket{\psi_{\mathrm{g}}}+e^{-i(E_{\mathrm{ex}}-\varepsilon)t}\ket{1}\ket{\psi_{\mathrm{ex}}}),
\label{Eq: after phase}
\end{aligned}
\end{equation}
where $E_{\mathrm{g}}$ and $E_{\mathrm{ex}}$ are the eigenvalues of the ground and excited states, respectively.
Suppose here the state before the measurement as
\begin{equation}
\begin{aligned}
\ket{\phi} = \frac{1}{\sqrt{2}}U_{\mathrm{prep}}^{\dagger}(e^{-iE_{\mathrm{g}}t}\ket{0}\ket{\psi_{\mathrm{g}}}+e^{-i(E_{\mathrm{ex}}-\varepsilon)t}\ket{1}\ket{\psi_{\mathrm{ex}}}). 
\label{Eq: state before the measurement}
\end{aligned}
\end{equation}
Then, the probability of having the measurement result $\Vec{x}=\Vec{0}$, i.e., the likelihood function $P(\Vec{0}| \varepsilon)$, is given by
\begin{equation}
\begin{aligned}
P(\Vec{0}| \varepsilon)
&=\Tr[(\ket{0}\bra{0})^{\otimes N+1} \ket{\phi}\bra{\phi}] \\
&\approx \frac{1}{4} |(\bra{0}\bra{\psi_{\mathrm{g}}}+\bra{1}\bra{\psi_{\mathrm{ex}}}) (e^{-iE_{\mathrm{g}}t}\ket{0}\ket{\psi_{\mathrm{g}}}+e^{-i(E_{\mathrm{ex}}-\varepsilon)t}\ket{1}\ket{\psi_{\mathrm{ex}}})|^2\\
&= \frac{1}{4}|e^{-iE_{\mathrm{g}}t}+e^{-i(E_{\mathrm{ex}}-\varepsilon)t}|^2\\
&= \frac{1}{2}(1+\cos(E_{\mathrm{gap}}-\varepsilon)t),
\label{Eq: after measurement}
\end{aligned}
\end{equation}
and the final expression can be approximated as
\begin{equation}
\begin{aligned}
P(\Vec{0}| \varepsilon) \approx {\mathrm{exp}}\left\{ -\frac{t^2}{4}\left( \varepsilon - E_\mathrm{gap} \right)^2 \right\},
\label{Eq: exp approximation after measurement}
\end{aligned}
\end{equation}
where $E_{\mathrm{gap}}$ is a energy gap represented as $E_{\mathrm{gap}}=E_{\mathrm{ex}}-E_{\mathrm{g}}$.
Therefore, by calculating the probability that all qubits are measured at 0 in an appropriate range of $\varepsilon$, the likelihood function can be approximated by a Gaussian distribution.

The number of two-qubit gates required to execute the circuit of Fig.~\ref{fig: Overview_3column.pdf}(c) is at most
\begin{equation}
    \begin{aligned}
    3\times\left\{N\times\ceil{\frac{d_{\mathrm{prep}}}{2}} \times 2 + (N-1) \times \ceil{\frac{d_{\mathrm{evol}}}{2}} \times \frac{t}{\Delta t}\right\},
    \label{Eq: the number of gate for the circuit execution}
    \end{aligned}
\end{equation}
where the first 3 comes from the fact that any two-qubit gate can be executed with up to 3 native two-qubit gates~\cite{Vatan2004-ck, Drury2008-cw}.
In the real device execution, the gate count would be smaller than the above value due to the transpilation and the error suppression, where optimization level 3 was adopted in the Qiskit transpilation.

Our phase difference estimation algorithm is implemented on Qiskit~\cite{Javadi-Abhari2024-qm} with the FireOpal error suppression module of Q-CTRL~\cite{Mundada2023-rl,Sachdeva2024-vr} by the Python language. 
FireOpal is a package of deterministic error-suppression workflow, and the overview of the execution procedure is as follows~\cite{Mundada2023-rl}; a user first create quantum circuits (as a quantum assembly language, QASM) and submit them to Q-CTRL through the FireOpal module. The module first executes the front-end compiler including mathematical reduction of the gate count of quantum circuits. Then it executes an error-reducing back-end compiler including an error-aware hardware mapping, an elimination of circuit crosstalk (dynamical decoupling), and an optimized gate replacement (e.g., AI-powered analog-level gate optimization). The created circuits are experimented on a real device. Before returning the results to the user, the module performs a measurement error mitigation including an AI-driven calibration routine.

We choose $m=21$, $R=10,000$, and $\Delta t = 0.1$. $\sigma^2_{\mathrm{init}}$ is $9.0$ in the result of Fig.~\ref{fig: Compare_1column.pdf} and is $4.0$ in the other results. $\mu_{\mathrm{init}}=0.0$ in most of the cases, but when the peak of the likelihood function diverged in the noiseless simulation, it deviated by 0.01.
In addition, in the molecular orbital model of hexatriene, the lack of precision in the approximation of $U_{\mathrm{evol}}$ sometimes caused the Gaussian fit to fail because the peak position deviated significantly from the value of the previous iteration.
In such cases, the value of $\varepsilon$ which has the maximum $p$ was assigned to $\mu_\mathrm{{prior}}$ and restarted.
The total time $t$ is chosen as $1.8/\sigma^2_{\mathrm{prior}}$.
For the noiseless simulation, the termination condition of our algorithm is achieving $\sigma^2_{\mathrm{post}} \leq 0.005$.
In the real device execution, the algorithm stopped around 4,000-8,000 two-qubit gates in a circuit before the suppression due to FireOpal (or IBM Quantum Platform) system errors, e.g., QASM upload size limit, before reaching the above conditions excluding octatetraene in Supporting Information~\ref{sec: Error analysis for the algorithm}, where we accidentally reached the termination condition in octatetraene because of noise contamination.
We also stopped iterations when the Gaussian peak was 0 or the Gaussian fit failed due to small signals and noise affections. 
The prior Gaussian distribution is used as the initial guess of the Gaussian fitting of the likelihood function.

\begin{figure}[th!]
 \centering
 \includegraphics[width=1\textwidth]{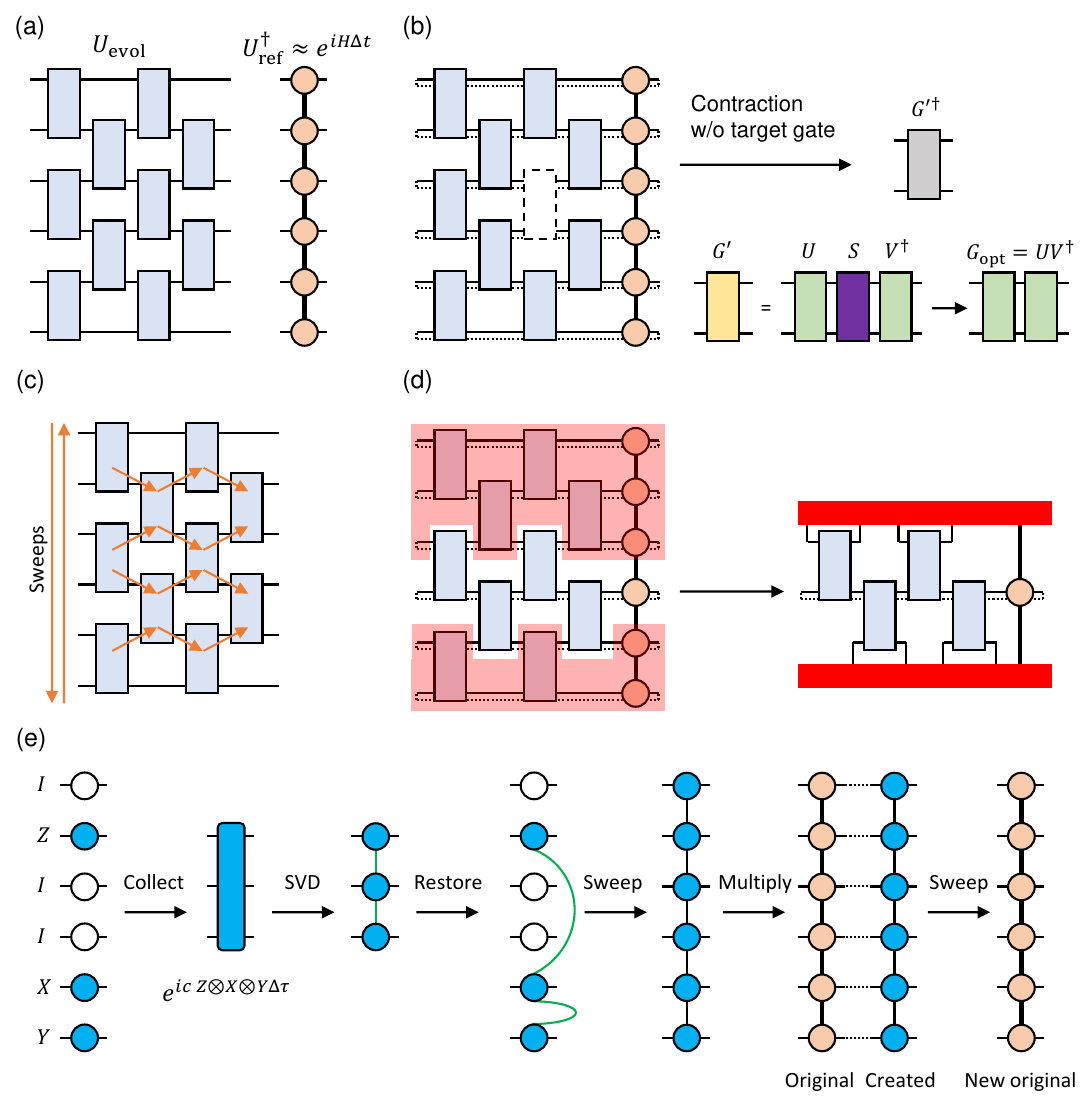}
\caption{Procedure of a unitary compression. Here is an example of six qubits and $d_{\mathrm{evol}}=4$. (a) Introduction of MPOs. Horizontal lines represent qubits. Vertical bold lines in $U_{\mathrm{ref}}^{\dagger}$ represent MPO virtual bonds. (b) Gate optimization. The dashed box represents a target gate to optimize. The dotted line connecting the left end and the right end represents a trace. (c) Optimization sequence. (d) Environmental tensors. (e) Example of creation of time evolution operator of $\exp{i c (I \otimes Z \otimes I \otimes I \otimes X \otimes Y) \Delta \tau}$.}
 \label{fig: UnitaryCompression_3column.pdf}
\end{figure}

\section{Unitary compression}
\label{sec: Unitary compression}
We approximate the time evolution in a single step $e^{-iH\Delta t}$ by $U_{\mathrm{evol}}$, where the error in the simulation for a total time $t$, $\norm{e^{-iHt}-U_{\mathrm{evol}}^{\frac{t}{\Delta t}}}_F$, is bounded as
\begin{equation}
\begin{aligned}
    \norm{e^{-iHt}-U_{\mathrm{evol}}^{\frac{t}{\Delta t}}}_F &= \norm{\sum_{k=1}^{t/\Delta t} e^{-iH (\frac{t}{\Delta t} - k)} (e^{-iH\Delta t}-U_{\mathrm{evol}}) U_{\mathrm{evol}}^{k-1} }_F \\
    &\leq \sum_{k=1}^{t/\Delta t}\norm{ e^{-iH (\frac{t}{\Delta t} - k)} (e^{-iH\Delta t}-U_{\mathrm{evol}}) U_{\mathrm{evol}}^{k-1} }_F \\
    &\leq \sum_{k=1}^{t/\Delta t} \norm{e^{-iH (\frac{t}{\Delta t} - k)}} \norm{e^{-iH\Delta t}-U_{\mathrm{evol}}}_F \norm{U_{\mathrm{evol}}^{k-1}}\\
    &= \frac{t}{\Delta t}\norm{e^{-iH\Delta t}-U_{\mathrm{evol}}}_F,
\label{Eq: upper bound of Frobenius norm}
\end{aligned}
\end{equation}
$\norm{\cdot}_F$ is the Frobenius norm, and $\norm{\cdot}$ is the operator norm. We used equality $A^n - B^n = \sum_{k=1}^n A^{n-k} (A-B) B^{k-1}$ in the second expression where $A$ and $B$ are matrices, and $n$ is an integer, triangle inequality in the third expression, and the inequalities $\norm{AB}_F \leq \norm{A}\norm{B}_F$ and $\norm{BA}_F \leq \norm{B}_F\norm{A}$ in the fourth expression.
The last inequalities can be shown as $\norm{AB}_F^2 = \Tr((AB)^{\dagger}AB) = \Tr(B^{\dagger}A^{\dagger}AB) \leq \lambda_{max}(A^{\dagger}A) \Tr(B^{\dagger}B) = \norm{A}^2\norm{B}_F^2$ where $\lambda_{max}(A^{\dagger}A)$ is the maximum eigenvalue of $A^{\dagger}A$, corresponding to the squared operator norm (and the squared maximum singular value) of $A$, and $A^{\dagger}A \leq \lambda_{max}(A^{\dagger}A) I^{\otimes N}$~\cite{Zou2020-vq}. $\norm{BA}_F \leq \norm{B}_F\norm{A}$ is also shown in almost the same procedure.

The unitary compression of $U_{\mathrm{evol}}$ is based on the procedure in Ref.~\cite{Causer2024-wd}.
We first prepare an MPO $U_{\mathrm{ref}}^{\dagger}$ corresponding to $e^{iH\Delta t}$ by using the second-order Trotter approximation for 100 sliced time steps, $\Delta t/100$.
Specifically, we start from an identity MPO and multiply by the time evolution MPO of the sliced step for each term of $H$, which is expanded by the tensor product of Pauli matrices.
To prevent divergence of the bond dimension, a sweep of singular value decomposition (SVD) truncating singular values below a threshold value of $10^{-12}$ was performed for the MPO each time applying the time evolution of the term.
Next, we optimize a brick-wall MPO $U_{\mathrm{evol}}$ to approximate $U_{\mathrm{ref}}$.
We consider minimizing the squared Frobenius norm of these operators,
\begin{equation}
\begin{aligned}
    \norm{ U_{\mathrm{ref}} - U_{\mathrm{evol}} }^2_F &= \Tr[U_{\mathrm{ref}}^{\dagger}U_{\mathrm{ref}}] + \Tr[U_{\mathrm{evol}}^{\dagger}U_{\mathrm{evol}}] - 2\Re\Tr[U_{\mathrm{ref}}^{\dagger}U_{\mathrm{evol}}]\\
    &= 2^{N+1}-2\Re\Tr[U_{\mathrm{ref}}^{\dagger}U_{\mathrm{evol}}].
\label{Eq: frobenius norm}
\end{aligned}
\end{equation}
The norm can be minimized by maximizing the $\Re\Tr[U_{\mathrm{ref}}^{\dagger}U_{\mathrm{evol}}]$ of the last term.

Figure~\ref{fig: UnitaryCompression_3column.pdf} shows the optimization procedure.
The orange tensor network represent $U_{\mathrm{ref}}^{\dagger}$, and the blue brick-wall tensor network represent $U_{\mathrm{evol}}$ in Fig.~\ref{fig: UnitaryCompression_3column.pdf}(a).
We explain how to optimize each of the gates in $U_{\mathrm{evol}}$;
$\Re\Tr[U_{\mathrm{ref}}^{\dagger}U_{\mathrm{evol}}]$ can be transformed as $\Re\Tr[(U_2 U_{\mathrm{ref}}^{\dagger}U_1) G]$ by representing $U_{\mathrm{evol}}=U_1 G U_2$ using the target gate $G$ and the the other gates $U_1$ and $U_2$. Then we consider finding the optimal $G$.
$\Re\Tr[(U_2 U_{\mathrm{ref}}^{\dagger}U_1) G]$ can be represented as $\Re\Tr[S( U^{\dagger}GV)]$ using the SVD $U_2 U_{\mathrm{ref}}^{\dagger}U_1 = V S U^{\dagger}$, where $U$ and $V$ are unitary matrices, and $S$ is a singular value (diagonal) matrix. 
$U^{\dagger} G V$ can be regarded as a unitary matrix, that is, each of the columns is a normalized vector, and thus $\Re\Tr[S( U^{\dagger}GV)] \leq \Re\Tr[S]$.
Therefore, when $U^{\dagger}GV$ is identity, i.e., $G = U V^{\dagger}$, $\Re\Tr[U_{\mathrm{ref}}^{\dagger}U_{\mathrm{evol}}]$ takes a maximum value.

To get the optimal gate in the actual implementation, we first contract all of the tensors except for the target gate and obtain a four-leg tensor $G'^{\dagger}$ in the left and right top panels of Fig.~\ref{fig: UnitaryCompression_3column.pdf}(b).
As in the right bottom panel of Fig.~\ref{fig: UnitaryCompression_3column.pdf}(b), the SVD is executed for $G'$, i.e., $G'=USV^{\dagger}$.
Then, we obtain the optimal gate $G_{\mathrm{opt}}=UV^{\dagger}$ by replacing $S$ to identity.
This optimization is performed for each gate.
As shown in Fig.~\ref{fig: UnitaryCompression_3column.pdf}(c), gate optimization is performed in the order of sweeping up and down the zigzag path from left to right.
The initial gate was prepared by QR decomposition of the identity operator plus a small random unitary operator.
The number of sweeps was 1,000 for the Hubbard models and 10,000 for linear polyenes.
To reduce the number of contractions, the upper and lower portions of the target gate are stored as an environmental tensor (red) and used as needed as in Fig.~\ref{fig: UnitaryCompression_3column.pdf}(d), and the cost of contracting the tensor network is $\order{\chi d_{\mathrm{evol}} 2^{4+2\ceil{d_{\mathrm{evol}}/2}}+\chi^2 2^{2+2\ceil{d_{\mathrm{evol}}/2}}}$, where $\chi$ is the maximum bond dimension in $U_{\mathrm{ref}}^{\dagger}$~\cite{Causer2024-wd}.
Note that in this study, when creating the $U_{\mathrm{ref}}^{\dagger}$ in Fig.~\ref{fig: UnitaryCompression_3column.pdf}(a), we reduce the computational cost by calculating the time evolution operators by excluding the identity operators from the tensor product of Pauli operators in a Hamiltonian.
Specifically, as in the example of Fig.~\ref{fig: UnitaryCompression_3column.pdf}(e), the following procedure is repeated for each term of the Hamiltonian; collecting only the Pauli $X, Y$, and $Z$ operators, performing time evolution operations, restoring them to their initial MPO sites using SVD, converting them to MPO by sweeping, and multiplying the created MPO by the original MPO to make a new original MPO. Thus, the cost of SVD is exponentially dependent on the locality of a Pauli term, rather than $N$. In addition, this exponential increase could be avoided by decomposing the collected exponential operator into local controlled-NOT and one-qubit gates.
The initial original MPO is identity, $c$ is a coefficient in the term, and $\Delta \tau = \Delta t/200$ in our case of second-order Trotter approximation.

To execute the unitary compression of $U_{\mathrm{prep}}$, we need to prepare $\ket{\mathrm{MPS}}=\frac{1}{\sqrt{2}}(\ket{0}\ket{\psi_{\mathrm{g}}}+\ket{1}\ket{\psi_{\mathrm{ex}}})$.
We calculate $\ket{\psi_{\mathrm{g}}}$ by the DMRG, and $\ket{\psi_{\mathrm{g}}}$ is an MPS as
\begin{equation}
\begin{aligned}
    \ket{\psi_{\mathrm{g}}} = \sum_{\Vec{a}}\sum_{\Vec{s}} A_{s_1 a_1} A_{s_2 a_1 a_2}\dots A_{s_N a_{N-1}}\ket{\Vec{s}},
    \label{Eq: MPS ground state only system}
\end{aligned}    
\end{equation}
where $\Vec{a}=a_1 a_2 \dots a_{N-1}$, $a_\iota \in \{0,1,\dots,\chi_{\iota}-1\}$ ($\iota = 1,2,\dots,N-1$) represents a virtual index, $\chi_{\iota}$ is a bond dimension, $\Vec{s}=s_1 s_2 \dots s_{N}$, and $s_\kappa \in \{0,1\}$ ($\kappa = 1,2,\dots,N$) represents a physical index.
$\ket{0}\ket{\psi_{\mathrm{g}}}$ can be represented as an MPS using a dummy tensor $A_{s_0}$,
\begin{equation}
\begin{aligned}
    \ket{0}\ket{\psi_{\mathrm{g}}} = \sum_{\Vec{a}}\sum_{s_0, \Vec{s}} A_{s_0} A_{s_1 a_1} A_{s_2 a_1 a_2}\dots A_{s_N a_{N-1}}\ket{s_0\Vec{s}},
\label{Eq: g for MPS}
\end{aligned}    
\end{equation}
where $A_{s_0}$ ($s_0 \in \{0,1\}$) is a tensor of $A_{0} = 1$ and $A_{1} = 0$.
$\ket{0}\ket{\psi_{\mathrm{ex}}}$ can also be expressed in an MPS by the same procedure as above,
\begin{equation}
\begin{aligned}
    \ket{1}\ket{\psi_{\mathrm{ex}}} = \sum_{\Vec{a}'}\sum_{s_0, \Vec{s}} A'_{s_0} A'_{s_1 a_1'} A'_{s_2 a_1' a_2'}\dots A'_{s_N a_{N-1}'}\ket{s_0\Vec{s}},
\label{Eq: ex for MPS}
\end{aligned}    
\end{equation}
where $\Vec{a}'$ is denoted in the same way as $\Vec{a}$ previously mentioned, and $A'_{s_0}$ is a tensor of $A'_{0} = 0$ and $A'_{1} = 1$.
We add the two states to prepare an unnormalized superposition state,
\begin{equation}
\begin{aligned}
    \ket{0}\ket{\psi_{\mathrm{g}}} + \ket{1}\ket{\psi_{\mathrm{ex}}} = \sum_{\tilde{a}_0\Vec{\tilde{a}}}\sum_{s_0, \Vec{s}} \tilde{A}_{s_0 \tilde{a}_0} \tilde{A}_{s_1 \tilde{a}_0 \tilde{a}_1} \tilde{A}_{s_2 \tilde{a}_1 \tilde{a}_2}\dots \tilde{A}_{s_N \tilde{a}_{N-1}}\ket{s_0\Vec{s}},
\label{Eq: superposition of MPS}
\end{aligned}    
\end{equation}
where
\begin{equation}
\begin{aligned}
&\tilde{A}_{s_0 \tilde{a}_0} =
\begin{cases}
    A_{s_0} & \text{if } \tilde{a}_0 = a_0 \\
    A'_{s_0} & \text{if } \tilde{a}_0 = a'_0
\end{cases}
\\
&\tilde{A}_{s_1 \tilde{a}_{\iota-1} \tilde{a}_{\iota}} =
\begin{cases}
    A_{s_\iota a_{\iota-1} a_{\iota}} & \text{if } \tilde{a}_{\iota-1} = a_{\iota-1}, \tilde{a}_{\iota} = a_{\iota} \\
    A'_{s_\iota a'_{\iota-1} a'_{\iota}} & \text{if } \tilde{a}_{\iota-1} = a'_{\iota-1}, \tilde{a}_{\iota} = a'_{\iota}\\
    0 & \text{otherwise}
\end{cases}
\\
&\tilde{A}_{s_N \tilde{a}_{N-1}} =
\begin{cases}
    A_{s_N a_{N-1}} & \text{if } \tilde{a}_{N-1} = a_{N-1} \\
    A'_{s_N a'_{N-1}} & \text{if } \tilde{a}_{N-1} = a'_{N-1}
\end{cases}
\label{Eq: enlarged tensor}
\end{aligned}    
\end{equation}
$\Vec{\tilde{a}} \in \{\Vec{a}, \Vec{a}'\}$ , and $\tilde{a}_0 \in \{a_0=0, a_0'=1\}$.
That is, the superposition state can be represented as an MPS.
$\ket{\mathrm{MPS}}$ is obtained by normalization and transformation to the left canonical form of the superposition state.

The optimization procedure of a brick-wall MPO $U_{\mathrm{prep}}$ to approximate $\ket{\mathrm{MPS}}$ is almost the same as $U_{\mathrm{evol}}$ since we can execute the same procedure by substituting $U_{\mathrm{evol}}$ to $U_{\mathrm{prep}}$ and $U_{\mathrm{ref}}^{\dagger}$ to $\ket{0}^{\otimes N+1}\bra{\mathrm{MPS}}$ in $\Re\Tr[U_{\mathrm{ref}}^{\dagger}U_{\mathrm{evol}}]$ of Eq.~\eqref{Eq: frobenius norm}, that is,
\begin{equation}
    \begin{aligned}
        \Re\Tr[\ket{0}^{\otimes N+1}\bra{\mathrm{MPS}} U_{\mathrm{prep}}] = \bra{\mathrm{MPS}} U_{\mathrm{prep}} \ket{0}^{\otimes N+1},
        \label{Eq: optimization cost expression for uprep}
    \end{aligned}
\end{equation}
which is the same as in Eq.~\eqref{Eq: prep metric}.
The number of sweeps was 1,000 for all the models.
Note that MPS can be embedded with sequential gates without optimization~\cite{Ran2020-li}, but it was not adopted in our case because of the increasing depth and idle time of each qubit.

The remaining calculation conditions are described as follows.
The unitary compression was implemented in ITensor~\cite{Fishman2020-lw} package of the Julia language.
Hamiltonian prepared by Qiskit (see Sec.~\ref{sec: model construction}) was converted to an ITensor format, and $U_{\mathrm{prep}}$ and $U_{\mathrm{evol}}$ obtained by unitary compressions were saved in NumPy~\cite{Harris2020-fh} format, and then the phase difference estimation was performed using Qiskit with the compressed gates.
The number of the DMRG sweeps when preparing $\ket{\mathrm{MPS}}$ was set to 20, the maximum bond dimension in SVD was set to 10 for the first 3 sweeps, 50 for the next 12 sweeps, and 1,000 for the last 5 sweeps, and the cutoff threshold in SVD was $10^{-12}$ for the Hubbard models and $10^{-8}$ for linear polyenes.
The DMRG of the excited states was executed to reduce overlap with the MPS of the ground state.
Note that in ITensor's code implementation, the MPS of Eq.~\eqref{Eq: superposition of MPS} is obtained simply by adding Eq.~\eqref{Eq: g for MPS} and Eq.~\eqref{Eq: ex for MPS}, i.e., and we did not need to construct the tensor of Eq.~\eqref{Eq: superposition of MPS} explicitly.

\section{Model construction}
\label{sec: model construction}
Qiskit~\cite{Javadi-Abhari2024-qm} was used for the fermionic model construction and Jordan--Wigner transformation for the Hubbard models and linear polyenes.
The electronic structures of linear polyenes were calculated by the PySCF package~\cite{Sun2020-sa}, and the geometry were optimized using the Hartree--Fock method with the 6-31G* basis set by the PySCF and geomeTRIC~\cite{Wang2016-pl} packages.
The molecular orbitals were localized by Boys localization~\cite{Foster1960-kw} for $\pi$ and $\pi^*$ orbitals. Fermionic Hamiltonians for hexatriene, octatetraene and decapentaene were constructed as 12-qubit CASCI (6e 6o), 16-qubit CASCI (8e 8o), and 20-qubit CASCI (10e 10o) problems, respectively.

The orbital order was optimized using a genetic algorithm implemented with the DEAP library~\cite{Fortin2012-fz}. The cost function is $C$ in Eq.~\eqref{eq: orbital sorting cost}, the parameters were set with a population size of 50, a crossover probability of 70\%, a mutation probability of 20\%, and 100 generations. Ordered crossover and index-shuffling mutation were applied, and tournament selection with a size of three was used.
Note that while we utilized the Jordan--Wigner transformation after the reordering, incorporating fermion-qubit mappings considering circuit topology~\cite{Tilly2022-pb, Miller2023-sd} are left for future work.

\begin{figure}[ht!]
 \centering
 \includegraphics[width=0.5\textwidth]{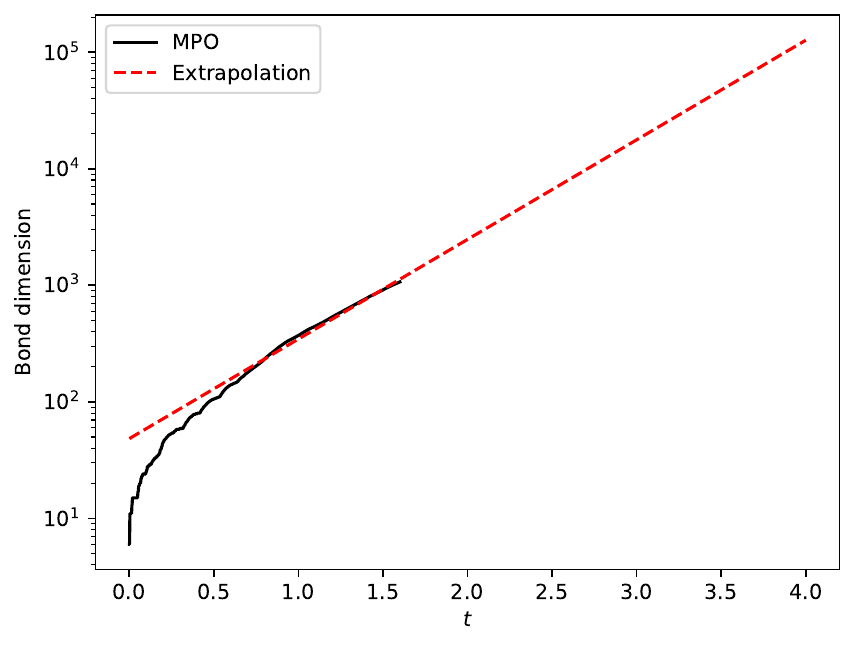}
\caption{Bond dimension of the time evolved MPO versus the time duration $t$.}
 \label{fig: plot_MPO_extrapolarion.pdf}
\end{figure}

\section{Time scale for the quantum advantage}
\label{sec: time scale for the quantum advantage}

The maximum time length taken in this study seems to be within the range where classical computers can simulate using the MPO framework. To estimate a classically-simulable regime, we additionally studied the time dependence of the bond dimension of MPO for the Hubbard model with 20 system qubits. The results are shown in Fig.~\ref{fig: plot_MPO_extrapolarion.pdf}. The MPO sliced time step width is the same as in this study, $\Delta t/100 = 10^{-3}$. We calculate it the bond dimension up to $t=1.6$ (black curve) and extrapolate the results by fitting it to an exponential function (red line). Considering that the limit of the bond dimension that can be executed in HPC is $10^{4}$-$10^{5}$~\cite{Ganahl2023-nh}, the classical calculation would become difficult in the timescale longer than $t\approx3.5$. In our demonstration, 20 qubit system real device execution was able to run up to $t=3.3$, so it appears that this naive comparison shows comparable performance. Therefore, a quantum advantage can be expected in the near future with improved device performance and algorithms. We finally note that by taking into account the tight cutoff threshold of SVD $10^{-12}$ and the approximation errors in the transformation from the MPO to a brick-wall circuit, minor modifications would lean the advantage to the classical side in the current situation.

\section{Error analysis for the algorithm}
\label{sec: Error analysis for the algorithm}

We note the error robustness in our algorithm.
Assume the global depolarization channel for the state before the measurement $\ket{\phi}$ in Eq.~\eqref{Eq: state before the measurement},
\begin{equation}
    \begin{aligned}
        \mathcal{E}(\ket{\phi}\bra{\phi}) = (1-p_{\mathrm{dep}})\ket{\phi}\bra{\phi} + p_{\mathrm{dep}}\frac{1}{2^{N+1}}I^{\otimes N+1},
        \label{Eq: error before measurement}
    \end{aligned}
\end{equation}
where $p_{\mathrm{dep}}$ is an error rate of the depolarization channel.
The probability in Eq.~\eqref{Eq: after measurement} becomes
\begin{equation}
    \begin{aligned}
        \Tr[(\ket{0}\bra{0})^{\otimes N+1} \mathcal{E}(\ket{\phi}\bra{\phi})] \approx (1-p_{\mathrm{dep}})\frac{1}{2}(1+\cos(E_{\mathrm{gap}}-\varepsilon)t) + \frac{p_{\mathrm{dep}}}{2^{N+1}},
        \label{Eq: error after measurement}
    \end{aligned}
\end{equation}
and the error does not change the peak position while the amplitude decreases.
In addition, the second term, which comes from the effect of depolarization noise, is exponentially decaying with respect to $N$.
When calculated with a similar setup in QPE [Fig.~\ref{fig: Overview_3column.pdf}(a)] and QPDE [Fig.~\ref{fig: Overview_3column.pdf}(b)], the second term becomes $p_{\mathrm{dep}}/2$, i.e., constant, due to the ancilla-only measurement~\cite{Wiebe2016-kr}.
Therefore, the exponentially small noise affection is a feature of our algorithm.

\begin{figure}[ht!]
 \centering
 \includegraphics[width=0.5\textwidth]{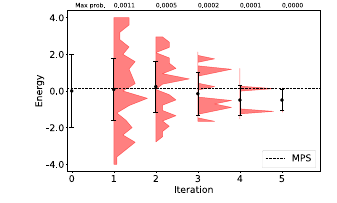}
\caption{Demonstration of our algorithm in the one-dimensional Hubbard models using the Heron device $ibm\_torino$ without the error suppression (21-qubit circuit). The circle and error bar denote the mean value and standard deviation, respectively, of the posterior distribution in the iteration. The sampled probabilities are shown in red, and each plot is normalized by the maximum value in each iteration, which is shown at the top of the plot.}
 \label{fig: WoQctrl_1column.pdf}
\end{figure}

\begin{figure}[ht!]
 \centering
 \includegraphics[width=1.0\textwidth]{{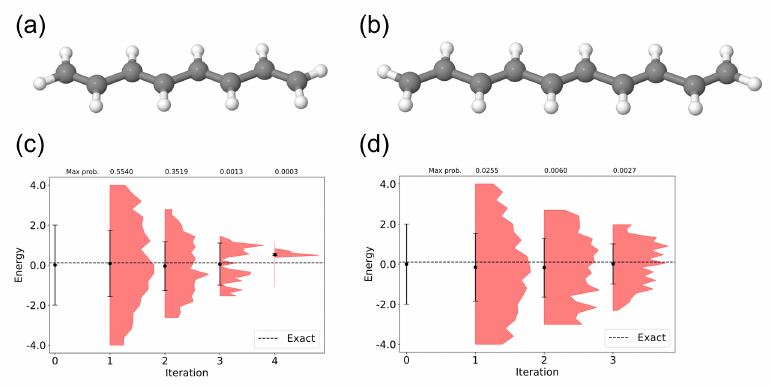}}
\caption{Demonstration of our algorithm in octatetraene (17-qubit circuit) and decapentaene (21-qubit circuit) using the Heron devices $ibm\_torino$ and $ibm\_fez$, respectively, with the error suppression. (a) the structure of octatetraene. (b) the structure of decapentaene. (c) the demonstration of our algorithm of octatetraene. The circle and error bar denote the mean value and standard deviation, respectively, of the posterior distribution in the iteration. The sampled probabilities are shown in red, and each plot is normalized by the maximum value in each iteration, which is shown at the top of the plot. (d) the demonstration of our algorithm of decapentaene.}
 \label{fig: Octatetraene_decapentaene_3column.pdf}
\end{figure}

The affection of the decay actually can be seen in the results of the real device execution.
For example, for the eight-qubit Hubbard model, the value of the second term of noise is  $2^{-9} \approx 2.0\times10^{-3}$ at maximum.
In fact, a uniform distribution of 0.001 magnitudes can be seen in the latter half of the iterations in Fig.~\ref{fig: Demonstration_3column.pdf}(b).
On the other hand, the values in 21- and 33-qubit circuits are $2^{-21} \approx 4.8\times10^{-7}$ and $2^{-33} \approx 1.2\times10^{-10}$, respectively, and the sharp peak can be seen of Fig.~\ref{fig: Demonstration_3column.pdf}(d) despite a small signal of $0.001\textendash{}0.01$  magnitude.
As related data, the 20 qubit Hubbard model without error suppression is shown in Fig.~\ref{fig: WoQctrl_1column.pdf}.
A small peak is seen at the beginning, but the signal disappears with the proceeding iteration.
In the final iteration, there is no signal at all, meaning that there is not only no signal but also no noise effect.
The same tendency was found in the octatetraene (17-qubit circuit) in Fig.~\ref{fig: Octatetraene_decapentaene_3column.pdf} (a) and (c); only a few shots were detected in the final iteration since the number of detection from the noise affection is estimated as $2^{-17} \times 10,000 \times 21 \approx1.6$ under 10,000 shots/circuit and 21 circuits/iteration, which led to a coincidentally sudden convergence in the final iteration. 
Here, $d_{\mathrm{prep}}=d_{\mathrm{evol}}=10$, the gap in the iteration of the final (before the final) is $0.525\pm0.051$ ($0.048\pm1.109$), the exact gap is 0.113, and the number of two-qubit gates is 4056/638 (1794/504).
We also executed decapentaene (21 qubit-circuit) in Fig.~\ref{fig: Octatetraene_decapentaene_3column.pdf} (b) and (d), and the gap is $0.009\pm1.005$, where $d_{\mathrm{prep}}=8, d_{\mathrm{evol}}=10$, the exact gap is 0.105, and a two-qubit gate count is 2730/693.
These results suggest that our algorithm can estimate the energy for large qubits as long as the signal is detected.

\begin{figure}[]
 \centering
 \includegraphics[width=1\textwidth]{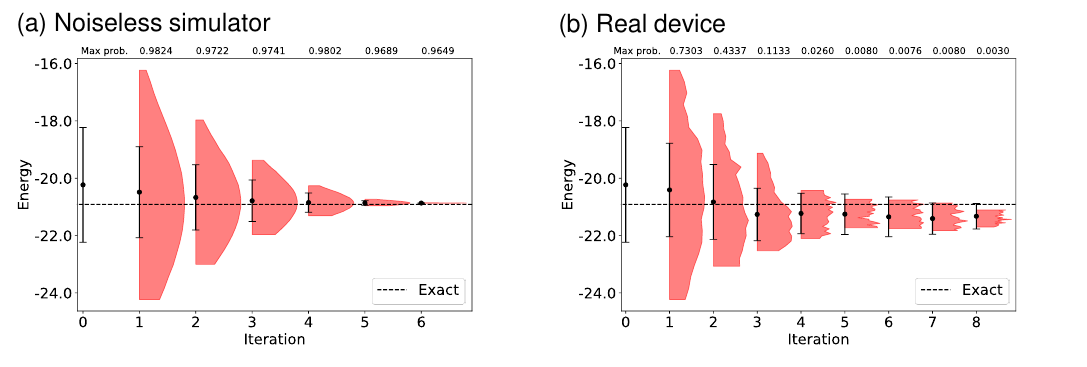}
\caption{Demonstration of our algorithm for the FCI calculation in the Hubbard model (nine-qubit circuit) using (a) the noiseless simulator and (b)  the Heron device $ibm\_torino$ with the error suppression. For ease of understanding, the signs of energy value are reversed. The circle and error bar denote the mean value and standard deviation,  respectively, of the posterior distribution in the iteration. The sampled probabilities are shown in red, and each plot is normalized by the maximum value in each iteration, which is shown at the top of the plot.}
 \label{fig: FCI_3column.pdf}
\end{figure}

\section{FCI energy calculation}
\label{sec: FCI energy calculation}
The FCI energy can be calculated by substituting $\ket{\psi_{\mathrm{ex}}}$ for a vacuum state $\ket{0}^{\otimes N}$ in the superposition state.
Since the vacuum state is trivially MPS and the energy of the vacuum state is zero, Eq.~\eqref{Eq: after measurement} becomes
\begin{equation}
\begin{aligned}
&\Tr[(\ket{0}\bra{0})^{\otimes N+1} \ket{\phi}\bra{\phi}] \approx \frac{1}{2}(1+\cos(E_{\mathrm{g}}+\varepsilon)t),
\label{Eq: fci estimation}
\end{aligned}
\end{equation}
the (sign reversed) ground state energy can be calculated in our algorithm.
Fig.~\ref{fig: FCI_3column.pdf} shows the demonstration of the eight-qubit Hubbard model, where $U_{\mathrm{evol}}$ was shared with the gap estimation in the main text, $U_{\mathrm{prep}}$ was prepared by $d_{\mathrm{prep}}=6$ with $f(U_{\mathrm{prep}};~\ket{\mathrm{MPS}}) = 0.99$, and $\mu_{\mathrm{init}}$ was set to the value when the DMRG is performed with the maximum bond dimension as two.
The results of the noiseless simulator in Fig.~\ref{fig: FCI_3column.pdf}(a) show that the distribution approaches the exact solution with each iteration. The final value is $-20.866\pm0.005$, where the exact solution is $-20.911$.
We also confirmed that the algorithm works on the real device (especially when the maximum probability in an iteration is greater than 0.01) in Fig.~\ref{fig: FCI_3column.pdf}(b), where the energy of a final iteration is $-21.326\pm0.444$.

\begin{figure}[ht!]
 \centering
 \includegraphics[width=1.0\textwidth]{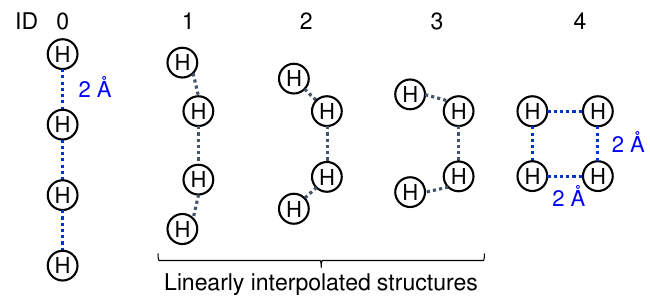}
\caption{Calculated hydrogen cluster models. The bond lengths between two hydrogen atoms in ID 0 and 4 are 2 Angstrom. The three structures are linearly interpolated between ID 0 and 4.}
 \label{fig: Hcluster_benchmark_1column.pdf}
\end{figure}

\begin{table}[ht!]
\centering
\caption{Results of hydrogen cluster models. The energy unit is Hartree. ``Gap (TQPDE)'' and ``Gap (exact)'' are the energy gaps of the tensor-based QPDE and exact, respectively. ``Error'' is the difference between Gap (TQPDE) and Gap (exact).}
\label{tab: Hcluster_benchmark_1column_result}
\begin{tabular}{llllll}
 \toprule
ID & $f(U_{\mathrm{prep}};~\ket{\mathrm{MPS}})$ & $\delta(U_{\mathrm{evol}};~U_{\mathrm{ref}})$ & Gap (TQPDE) & Gap (exact) & Error \\ \hline \hline
0 & 1.00 & 0.0055 & 0.122 & 0.0159 & 0.1057 \\
1 & 1.00 & 0.0068 & 0.200 & 0.0793 & 0.1212 \\
2 & 1.00 & 0.0074 & 0.276 & 0.1283 & 0.1481 \\
3 & 1.00 & 0.0076 & 0.218 & 0.0769 & 0.1406 \\
4 & 0.92 & 0.0137 & -0.028 & 0.0171 & 0.0448 \\
 \hline 
\end{tabular}
\end{table}

\section{Additional benchmarks for interaction complexity}
\label{sec: additional benchmarks for interaction complexity}
We performed our algorithm on five structures of $\mathrm{H_4}$ clusters that gradually changed from linear to square structures using noiseless simulations.
We show the result in Table~\ref{tab: Hcluster_benchmark_1column_result}, where $d_{\mathrm{prep}} = d_{\mathrm{evol}} = 8$. We found that all of the models achieved a level of accuracy for the error similar to that of the hexatriene. However, from the metrics of MPS and MPO, $f(U_{\mathrm{prep}};~\ket{\mathrm{MPS}})$ and $\delta(U_{\mathrm{evol}};~U_{\mathrm{ref}})$, respectively, we found that as the interaction changed from one-dimensional to two-dimensional, the accuracy tended to decrease. Therefore, although our approach can be applied to complex interactions, it was found to be more effective for systems that are close to one-dimensional.

\bibliographystyle{plain}
\bibliography{TQPDE}

\end{document}